\documentclass[12pt]{article}
\usepackage{amsmath}
\usepackage{epsfig}
\usepackage{amstext,amssymb,amsfonts}

\advance\voffset by -2.0cm
\advance\hoffset by -1.75cm
\def\Tr{\,{\rm Tr}\, }

\def\be{\begin{equation}}
\def\ee{\end{equation}}
\def\ba{\begin{eqnarray}}
\def\ea{\end{eqnarray}}

\newcommand{\A}{{\cal A}}

\renewcommand{\H}{{\cal H}}

\newcommand{\ket}[1]{|{#1}\rangle}

\newcommand{\ie}{{\it i.e.~}}

\begin{document}
\vspace*{1.5cm}

\begin{center}
{\Large 
{\bf Modular differential equations for} 
\vspace{0.5cm}

{\bf  torus one-point functions}}
\vspace{2.5cm}

{\large Matthias R.\ Gaberdiel$^{1}$}
\footnotetext{$^{1}${\tt E-mail: gaberdiel@itp.phys.ethz.ch}} 
{\large and} 
{\large Samuel Lang$^{2}$}
\footnotetext{$^{2}${\tt E-mail: slang@student.ethz.ch}} 
\vspace*{0.5cm}

Institut f{\"u}r Theoretische Physik, ETH Z{\"u}rich\\
CH-8093 Z{\"u}rich, Switzerland\\
\vspace*{3cm}

{\bf Abstract}
\end{center}

It is shown that in a rational conformal 
field theory every torus one-point function 
of a given highest weight state 
satisfies a modular differential equation. 
We derive and solve these differential
equations explicitly for some 
Virasoro minimal models. In general, however,
the resulting amplitudes do not seem to be 
expressible in terms of standard transcendental functions.


\newpage
\renewcommand{\theequation}{\arabic{section}.\arabic{equation}}


\section{Introduction}
\setcounter{equation}{0}

It has been known for some time that every rational conformal field
theory satisfies a common modular differential equation that is solved
by all characters of the (finitely many) representations.  This fact
was first observed, using the transformation properties of the
characters under the modular group, in
\cite{Eguchi:1986sb,Anderson:1987ge,MMS,MMS1}; later developments of
these ideas are described in
\cite{Kiritsis:1989kc,Eholzer,ES}. Following the work of Zhu
\cite{Zhu}, the modular transformation properties of the characters
were derived from first principles (see also
\cite{Nahm:1991ie}). Zhu's derivation suggests that the modular
differential equation is a consequence of a null-vector relation in
the vacuum Verma module \cite{Gaberdiel:2007ve,GaKe}, see also
\cite{Flohr:2005cm}.

This modern point of view suggests that not only the characters
of a rational conformal field theory are characterised by a modular
differential equation, but that the same is also true for the torus one-point
functions of a given highest weight state \cite{Mi} (see also \cite{Kiritsis:1989kc}). 
Indeed, every  highest weight state of a rational conformal field theory has a non-trivial 
null-vector which in turn leads to a modular differential equation
for the associated torus one-point functions. In this paper we 
explain how to derive this differential equation, and then
exemplify this general method with the case of the Virasoro
minimal models. In particular, we determine closed form
expressions for the torus one-point functions of all highest weight
states of the Yang-Lee, the Ising and the tricritical Ising model;
for the case of the Ising model our answer reproduces the 
results of \cite{Francesco,Mi}. We also make some general statements about  
the one-point functions of the $(1,4)$, $(4,1)$ and $(2,2)$ field of 
a general minimal model. As we explain, generically the resulting
amplitudes cannot be expressed in terms of standard 
transcendental functions. 

For affine theories, torus one-point functions have been studied
before, using the generalisation of the Knizhnik-Zamolodchikov
equation to surfaces of higher genus \cite{Bernard:1987df}, see also 
\cite{Felder:1994gk}. The analysis of Zhu \cite{Zhu} was generalised to 
the case of torus one-point functions in \cite{Mi}, see also 
\cite{Huang:2003yq}; our derivation of the modular differential equation
is a direct application of this analysis. The modular covariance properties
of the torus one-point functions were derived in \cite{Felder} 
for the case of the minimal models, and in \cite{Mi} for general
rational conformal field theories. More recently, these torus
amplitudes have been studied from the point of view of the 
representation of the modular group they give rise to \cite{Terry1,Terry2}. 
\smallskip

The paper is organised as follows. In the following section we
show how to derive the modular differential equation for the 
torus one-point functions. The question of whether the 
resulting functions can be expressed in terms of standard transcendental
functions is addressed in section 2.1. Section 3 applies these
general ideas to the Virasoro minimal models. Finally, section 4
contains our conclusions. There are three appendices where 
our conventions and a technical lemma (due to Terry Gannon)
is described.

\section{The differential equation}
\label{method}

\setcounter{equation}{0}

Suppose that ${\cal A}$ is a rational chiral algebra (or vertex operator
algebra); for a brief introduction to our conventions see appendix~A. 
Since $\mathcal{A}$ is rational, it has only finitely many inequivalent 
highest weight representations; we shall denote these by $\H_j$. Let $v$ be 
a highest weight state of a representation $v\in\H_j$. We want to study the 
torus one-point functions of $v$, \ie
\be
\Tr_{\H_l} \Bigl( V(v,z) \, q^{L_0-\frac{c}{24}} \Bigr) \ .
\ee
Here $\H_l$ is any representation of the chiral algebra
$\A$. Obviously, in order for this torus amplitude to be non-trivial
we need that the fusion rules allow for the fusion 
\be\label{fu}
\H_l \otimes \H_j \supset \H_l \ . 
\ee
Since 
\be
{}[L_0,V(v,z)] = \left(z \frac{d}{dz} + h \right) V(v,z) \ , 
\qquad \hbox{with} \quad
L_0 v = h v 
\ee
it follows that the torus amplitude has a 
trivial $z$-dependence; indeed, if we define 
\be
\chi_l(v;\tau) = z^{h} 
\Tr_{\H_l} \Bigl( V(v,z)\, q^{L_0-\frac{c}{24}} \Bigr)  =
\Tr_{\H_l} \Bigl( V(z^{L_0} v,z)\, q^{L_0-\frac{c}{24}} \Bigr) 
\ ,
\ee
then $\chi_l(v,\tau)$ is 
in fact independent of $z$. Note that the last expression is even 
defined for $v$ that are not eigenvectors of $L_0$. In the following
we shall sometimes set $z=1$ in order to simplify our expressions.

The idea of our analysis is to derive a
differential equation in $\tau$ (that is independent of which representation
$\H_l$ is considered, but does depend on $v$) for
these amplitudes. Following the analysis of \cite{Zhu,Mi},
this can be done in essentially the same way as
in \cite{GaKe}. The key step of the argument is
the recursion relation 
\be\label{a}
{}\mbox{Tr}_{\mathcal{H}_j} 
\left(V(a_{[-h_a]}v,1)q^{L_0}\right)
=  \mbox{Tr}_{\mathcal{H}_j}\left(o(a)V(v,1)q^{L_0}\right)
 + \sum_{k=1}^{\infty}
G_{2k}(q)\, \mbox{Tr}_{\mathcal{H}_j}
\left(V(a_{[2k-h_a]}v,1)q^{L_0}\right)\ , 
\ee
where $G_n(q)$ is the $n^{th}$ Eisenstein series that is defined in 
appendix \ref{eisenstein}. Note that the only difference to
\cite{GaKe} is that $v$ is now not necessarily an element of the
chiral algebra ${\cal A}$. However, the commutation relations of $a_n$
with $V(v,z)$ --- these are the main ingredient in the derivation ---
are the same, independent of whether $v$ is in the chiral algebra or
not, and thus the argument goes through without any change, see 
also \cite{Mi}.

If we replace $a$ by $L_{[-1]}a$ in
(\ref{a}) and use that $\left(L_{[-1]}a\right)_{[n]}=-(n+h_a)a_{[n]}$, 
as well as $o(L_{[-1]}a)=(2 \pi i)o(L_{-1}a+L_0a)=0$, we get 
\be\label{1}
\mbox{Tr}_{\mathcal{H}_j} 
\left(V(a_{[-h_a-1]}v,1)q^{L_0}\right)
+ \sum_{k=1}^{\infty}(2k-1)\,  G_{2k}(q)\, 
\mbox{Tr}_{\mathcal{H}_j} \left(V(a_{[2k-h_a-1]}v,1)q^{L_0}\right)=0\ .
\ee
Actually the term with $k=1$ does not contribute since it is
commutator (see \cite{Zhu,Mi,GaKe})
\be
[o(a),V(v,z)] = V(a_{[-h_a+1]}v,z)
\ee
that vanishes in the trace. 

Given these observations we now make the following
definition. Let $\mathcal{H}_j[G_4(q), G_6(q)]$ denote the space of
polynomials in the Eisenstein series with coefficients in
$\mathcal{H}_j$. We then define $O_q(\mathcal{H}_j)$ to be the subspace of
$\mathcal{H}_j[G_4(q), G_6(q)]$ generated by the states of the form 
\begin{equation}\label{2}
O_q(\mathcal{H}_j): \quad 
a_{[-h_a-1]}v+\sum_{k=2}^{\infty}(2k-1)G_{2k}(q)a_{[2k-h_a-1]}v\ ,
\end{equation}
where $a\in {\cal A}$ and $v\in\mathcal{H}_j$. 
Note that the sum is finite, since $a_{[n]}$ annihilates $v$ for 
sufficiently large $n$.  It then follows from (\ref{1}) that 
\begin{equation}\label{3}
\chi_l(v;\tau)=0\ , \qquad \mbox{if } v\in O_q(\mathcal{H}_j)\ .
\end{equation}
Finally, we observe that
\begin{equation}\label{444}
a_{[-h_a-n]}v-(-1)^n\sum_{2k\geq n+1}
\binom{2k-1}{n}G_{2k}(q)a_{[2k-h_a-n]}v\in O_q(\mathcal{H}_j)\ , 
\quad n\geq 1\ ,
\end{equation}
as follows from (\ref{2}) by repeatedly replacing $a$ by $L_{[-1]}a$. 

With these preparations we can now construct a modular differential
equation for the torus one-point amplitudes. Suppose that 
$v\in \mathcal{H}_j$ is a highest weight state with conformal weight $h$. 
If the theory is rational, then it follows from the argument of
\cite{Abe} (together with the usual argument that is due to Zhu
\cite{Zhu}) that we can find an integer $s$ such that 
\begin{equation}\label{4}
(L_{[-2]})^s v + \sum_{r=0}^{s-2}g_r(q) (L_{[-2]})^r v \in
O_q(\mathcal{H}_j) \ , 
\end{equation}
where the $g_r(q)$ are modular forms of weight $2(s-r)$.  This then
implies that the one-point functions $\chi_l(v;\tau)$ satisfy a
modular differential equation of the form
\begin{equation}\label{5}
\left[D^{s,h}
+\sum_{r=0}^{s-2}f_r(q)D^{r,h} \right]
\chi_l(v;\tau)=0\ ,
\end{equation}
where $D^{t,h}$ is the order $t$ differential operator 
\begin{displaymath}
D^{t,h}=D_{2t-2+h}D_{2t-4+h}\dots D_h\ , 
\quad \mbox{with}\quad 
D_a=q\frac{d}{dq} - \frac{a}{4\pi^2}G_2(q)
=q\frac{d}{dq} - \frac{a}{12}E_2(q)\ ,
\end{displaymath}
and the $f_r(q)$ are modular forms of weight $2(s-r)$. To prove this
we first note that because of (\ref{3}) the right hand side of
(\ref{4}) vanishes inside the trace. On the other hand, using
(\ref{a}) repeatedly, each term on the left hand side can be written
as 
\begin{equation}\label{6}
 \mbox{Tr}_{\mathcal{H}_l}
\left(V((L_{[-2]})^r v,1)\, q^{L_0-\frac{c}{24}}\right)
=P_r^{(h)}(D)\, \chi_l(v;\tau)\ ,
\end{equation}
involving a modular covariant differential operator 
$P_r^{(h)}$ of order $r$. To see this, we note that 
for $r=1$ one gets, using (\ref{a}) and replacing 
$L_0$ by $L_0-\frac{c}{24}$,  
\begin{eqnarray}
\mbox{Tr}_{\mathcal{H}_l}\left(V(L_{[-2]}v,1)
q^{L_0-\frac{c}{24}}\right)
& = & (2\pi i)^2\,  
\mbox{Tr}_{\mathcal{H}_l}
\left( (L_0 - \frac{c}{24}) V(v,1)q^{L_0-\frac{c}{24}}\right)\nonumber \\ 
&& \quad +  h\, G_2(q)\mbox{Tr}_{\mathcal{H}_l} 
\left(V(v,1)q^{L_0-\frac{c}{24}} \right)\nonumber \\ 
& = & 
(2\pi i)^2 D_h \chi_l(v;\tau) \ ,
\end{eqnarray}
which is modular covariant with weight $2+h$ (since $\chi_l(v;\tau)$
is modular covariant with weight $h$). The case for general $r$
follows by again applying (\ref{a}), and using the same recursive
argument as in \cite{GaKe}. For the first few values of $r$, explicit
formulae for the operators $P_r^{(h)}$ are given in
appendix~\ref{proof}.

\subsection{Admissibility}

On general grounds we expect the different solutions of this
differential equation to correspond to the different torus one-point
amplitudes with an insertion of $v\in \H_j$. These different solutions
will be mapped into one another under the modular group; thus the
family of solutions forms a vector-valued (generalised) modular form
of weight $h$ \cite{Felder,Mi}. In fact, this property also follows
from the modular covariance of the differential equation derived
above. In order to account for the modular weight it is convenient to
make the ansatz  
\be\label{ana1}
\chi_l(v;\tau) = \eta(q)^{2h}\, g_l(q)\ , \qquad 
\eta(q) = q^{\frac{1}{24}} \prod_{n=1}^{\infty} (1-q^n) \ ,
\ee
where $\eta(q)$ is the Dedekind eta function.
Since it has modular
weight $1/2$, the $g_l(q)$ are then components of a vector-valued
modular function $\mathbb{X}: \mathbb{H}\rightarrow \mathbb{C}^r$
(where $\mathbb{H}$ is the upper half-plane and $r$ is the number
of different solutions $g_l$), satisfying the
transformation property 
\be
\mathbb{X}\left(\frac{a\tau +b}{c\tau +d}\right)
=\rho\left({a\,\,b}\atop {c\,\,d}\right)\mathbb{X}(\tau)\ .
\ee
Here $\left({a\,\,b} \atop {c\,\,d}\right)\in 
\mbox{SL}_2(\mathbb{Z})$ is an arbitrary group element, and 
$ \rho :\mbox{SL}_2(\mathbb{Z}) \rightarrow \mbox{GL}(r,\mathbb{C})$   
is a representation of the modular group. 

As we shall see, in certain simple examples we shall be able to give
very explicit formulae for the $g_l(q)$. However, even among the 
minimal models, this will not be possible in general. In fact, it 
is known \cite{Lang,Terry1} that the vector valued modular functions
can be expressed using known transcendental functions (in particular 
the Fricke functions) if the
representation $\rho$ of the modular group $\mbox{SL}_2(\mathbb{Z})$
is \textit{admissible}. Here admissible means that  
\begin{enumerate}
    \item $\Gamma(N)\subset$ ker$\rho$, for some integer $N$;
    \item $T$ is diagonal and $S^2$ is a permutation matrix.
\end{enumerate}
The subgroup $\Gamma(N)$ is defined by 
\be
\Gamma(N)=
\left\{\begin{pmatrix}a&b\\c&d\end{pmatrix}\in 
\mbox{SL}_2(\mathbb{Z}),\,|\,a,d\equiv 1\, 
(\mbox{mod }N) \mbox{ and }b,c\equiv 0 \,
(\mbox{mod }N)\right\}
\ee
and 
\be\label{Tdef}
T=\rho\begin{pmatrix}1&1\\0&1\end{pmatrix} \ , \qquad 
S=\rho\begin{pmatrix}0&-1\\1&0\end{pmatrix} \ .
\ee
It is often not easy to determine whether a representation
is admissible or not. However, there exists the following simple test
that we shall use below: Let $N$ be the order of the
modular $T$-matrix and let $t_1,\dots,t_n$ be the eigenvalues of $T$. 
If one can find an integer $m$ coprime to $N$, $1<m<N$, such that
the collection $t_1^{m^2},\dots,t_n^{m^2}$ does not agree 
(including multiplicities) with the original
collection of eigenvalues $t_1,\dots,t_n$, then the
representation is not admissible. The criterion, as well as its proof,
is due to Terry Gannon (see also \cite{Terry3}); more details
are given in appendix~\ref{proof}.

\section{Examples: Minimal models}
\setcounter{equation}{0}

We would now like to illustrate the results of the previous chapter
with some examples. In the following we shall concentrate on the
minimal models. Recall that the central charge of the $(p,q)$ 
minimal model is  \cite{DMS}
\be
c_{p,q} = 1 - \frac{6(p-q)^2}{pq} \ , 
\ee
and that the allowed representations are described by $(r,s)$ where
$r$ and $s$ are integers satisfying $1\leq r \leq q-1$ and 
$1\leq s \leq p-1$. The highest weight of the representation
corresponding to $(r,s)$ has conformal dimension 
\be
h_{r,s} = \frac{(pr-qs)^2-(p-q)^2}{4pq} \ ,
\ee
and we have the identifications $(r,s) \simeq (q-r,p-s)$. 
The representation labelled by $(r,s)$ has two independent null vectors:
one at level $rs$, and one at level $(q-r)(p-s)$.

\subsection{Simple examples: Yang-Lee and Ising}

It is instructive to start by analysing two simple examples. 

\subsubsection{Yang-Lee model}

The Yang-Lee model is the minimal model with $(p,q)=(5,2)$. Its
central charge is $c=-\frac{22}{5}$, and it has two highest-weight
representations with conformal weights
\be
h_{1,1}=h_{1,4}=0 \ , \qquad
h_{1,2}=h_{1,3}=-\frac{1}{5} \ . 
\ee
The only non-trivial representation is thus $\H_j=\H_{-\frac{1}{5}}$, and
since its Kac label is $(1,2)$, it has a null-vector at level $2$
\begin{equation}
\mathcal{N}_2=\left(L_{[-2]} -\frac{5}{2}
(L_{[-1]})^2\right)\ket{-1/5}\ .
\end{equation}
In the torus amplitude $L_{[-1]}$ descendants do not contribute since 
$o(L_{[-1]}b)=0$; thus the above null-vector leads simply to the
differential equation 
\begin{equation}\label{9}
D_{-1/5}\, \chi(-1/5;\tau) = 
\left[q\frac{d}{dq}+\frac{1}{60}E_2(q)\right]
\chi(-1/5;\tau) = 0 \ .
\end{equation}
It is not difficult to show that (\ref{9}) is solved by
\begin{equation}
\chi(-1/5;\tau)=\eta^{-2/5}(q)\ ,
\end{equation}
as follows from the well known relation 
\begin{equation}\label{10}
-4\pi i \frac{d}{d\tau}\ln(\eta(\tau))=G_2(\tau) 
= \frac{\pi^2}{3} \, E_2(\tau)\ .
\end{equation}
Note that the leading power of $\chi(-1/5;\tau)=\eta^{-2/5}(q)$ is 
\be
\eta^{-2/5}(q) = q^{-\frac{1}{60}} \prod_{n=1}^{\infty}
(1-q^n)^{-2/5} \ .
\ee
Thus the representation $\H_l$ in which the trace is taken 
is the $\H_{-1/5}$ representation since 
$-1/5-c/24 = -1/5 + 11/60=-1/60$. This is also compatible with
the fusion rules since 
\be
\left(-\frac{1}{5} \right)\otimes \left(-\frac{1}{5}\right) = 
\left( 0\right) \oplus \left( -\frac{1}{5} \right) 
\ee
contains the $(-1/5)$ representation. 
[On the other hand, the fusion of $(0)\otimes (-1/5)=(-1/5)$,
and hence does not contain $(0)$.]
\smallskip

Obviously, the representation with conformal weight $h_{1,3}=-1/5$
also has a null-vector at level $3$, which is explicitly given by
\be
{\cal N}_3 = \Bigl( L_{[-3]} - \frac{10}{9} L_{[-1]} L_{[-2]} 
+ \frac{25}{36} L_{[-1]} L_{[-1]} L_{[-1]} \Bigr) \ket{-1/5} \ .
\ee
However, this null vector leads to a trivial differential equation:
since $L_{[-1]}$ descendants do not contribute inside the trace, only
the first term can be non-trivial. However, it follows for example
from (\ref{1}) that it also vanishes inside any trace.

\subsubsection{Ising model}
The Ising model is the minimal model $(p,q)=(4,3)$ with 
central charge $c=\frac{1}{2}$. Its highest weight representations
have conformal weights 
\be
h_{1,1}=h_{2,3}=0 \ , \qquad
h_{1,2}=h_{2,2}=\frac{1}{16} \ , \qquad
h_{1,3}=h_{2,1}=\frac{1}{2} \ .
\ee
Let us first consider the one-point functions of the $h=\tfrac{1}{2}$
field. It has a null-vector at level $2$ which is of the form 
\be
\mathcal{N}_2=\left(L_{[-2]} - \frac{3}{4}(L_{[-1]})^2 \right) \ket{1/2}\ .
\ee
By the same arguments as above this leads to the differential
equation 
\be
D_{1/2}\, \chi(1/2;\tau) = 
\left[q\frac{d}{dq}-\frac{1}{24}E_2(q)\right]
\chi(1/2;\tau) =0 \ ,
\ee
whose unique solution is (see also \cite{Francesco,Mi})
\begin{equation}\label{11}
\chi(1/2;\tau)=\eta(q)\ .
\end{equation} 
Note that the leading exponent is 
$1/24 = 1/16-c/24 = 1/16 - 1/48$, and hence the character is taken in
the $\H_l=\H_{1/16}$ representation. This ties in with the fact that
there is the fusion rule
\be
\left(\frac{1}{2}\right)\otimes
\left(\frac{1}{16}\right) = \left(\frac{1}{16}\right) \ .
\ee
Actually  this is the only non-trivial torus one-point function since 
the fusion of $(1/2)$ with $(1/2)$ does not contain $(1/2)$. 
We also note in passing that the 
level three null vector of the $h=1/2$ representation leads again to a trivial
differential equation.
\smallskip

The situation is more interesting for the $h=1/16$ field, for which we
get two non-trivial differential equations, one from the null vector
at level two, and one from the null vector at level four. The first
one is simply  
\be
D_{1/16\, }\chi(1/16;\tau) = \left[ q \frac{d}{dq} 
- \frac{1}{192} E_2(q) \right] 
\chi(1/16;\tau)  = 0 \ ,
\ee
while the second one turns out to be (compare with the calculations of section 
\ref{level4})
\be
\left[D_{33/16} D_{1/16}-\frac{5}{576}E_4(q)\right]\chi(1/16;\tau) = 0\ .
\ee
One easily shows that there is no non-trivial solution to both of
these equations; this ties in with the fact that the fusion rules do
not allow for a non-trivial one-point function for the $h=1/16$ field
since 
\begin{equation}
\left(\frac{1}{16}\right)\otimes \mathcal{H}_i\not\supset \mathcal{H}_i\ ,
\qquad \mbox{for any highest-weight representation }\mathcal{H}_i \ .
\end{equation}

\subsection{A more general analysis}

We can now put these considerations into a somewhat more general
context. Each representation $(r,s)$ has 
two independent null-vectors at levels $N_1=rs$ and 
$N_2=(q-r)(p-s)$.  Note that $N_1$ is only odd if 
both $r$ and $s$ are odd; in this case, $N_2$ is 
necessarily even since $p$ and $q$ are coprime, and hence
cannot both be even. Thus either $N_1$ or $N_2$, or both
numbers are even. Without loss of generality, we may therefore
assume that $N_1=rs$ is even. 
\smallskip

As we have seen above, for each null-vector of even level $N$
we get a non-trivial modular differential equation of order $N/2$, while
a null-vector at odd level does not give rise to any non-trivial
constraint. (It  
is not difficult to show this in general since the recursion relations only
relate states at odd level to states at odd level.) From
what we have just explained, we therefore always have at least
one non-trivial differential equation of order $rs/2$. If in addition
$(q-r) (p-s)$ is even, then we have a second 
(linearly independent) differential equation. Since both equations
are linear differential equations in the same variable $\tau$, we then
do not in general expect to find any non-trivial solution (as was 
for example the case for the $h=1/16$ field in the Ising model). 
In fact, the absence of a solution can also be understood in general
using the constraints coming from the fusion rules (\ref{fu}).

To see this we first observe that $N_1$ and $N_2$ can only both
be even if (i) both $r$ and $s$ are even, and either $p$ or $q$ is even;
(ii) $r$ is even, $s$ is odd, and $p$ is odd; (iii) $s$ is even,
$r$ is odd, and $q$ is odd. Next we recall that the fusion of a 
representation $(l,m)$ with $(r,s)$ is given by \cite{DMS}
\be\label{fr}
(l,m)\otimes (r,s) = 
\bigoplus_{l'=|l-r|+1}^{{\rm min}(l+r-1,2q-l-r-1)}\,\, 
\bigoplus_{m'=|m-s|+1}^{{\rm min}(m+s-1,2p-m-s-1)} 
(l',m') \ , 
\ee
where in each sum $l'$ and $m'$ take only every other integer value.
It is then easy to see that in all three cases (i)--(iii), $(l,m)$ does not 
appear (up to the field identification $(l',m')\sim (q-l',p-m')$) in 
this fusion product. In particular, this then implies that there should
not exist any torus one-point function, since the condition (\ref{fu})
is not satisfied. 

For example, in case (i), the $l'$ and $m'$ that 
appear in (\ref{fr}) have the opposite cardinality from $l$ and $m$
respectively; thus we need to apply the field identification, but since
either $p$ or $q$ is even, this does not alter the cardinality of one of them.
In case (ii), we need to apply a field identification to the right-hand
side of (\ref{fr}) since $l'$ has the opposite cardinality from $l$. 
However, since $p$ is odd $p-m'$ then has the opposite cardinality from 
$m$. The analysis in case (iii) is identical.
\medskip

Thus we shall concentrate on the case that $N_1$ is even and 
$N_2$ is odd. This is the case if (a) both $r$ and $s$ are even, and
both $p$ and $q$ are odd; (b) $r$ is even and $s$ is odd, and 
$p$ is even and $q$ is odd; (c) $r$ is odd and $s$ is even, and 
$p$ is odd and $q$ is even. 

In each of these cases, we have one modular differential equation of 
order $rs/2$. We should thus expect that the $rs/2$ solutions correspond
to the different torus one-point amplitudes that are allowed by
the fusion rule condition (\ref{fu}). To see this we observe that the
fusion of $(l,m)$ with $(r,s)$ contains only representations of the
form (depending on the values of $(l,m)$ and $(r,s)$ some of these terms
may in fact not appear)
\be
(l+\Delta l, m+\Delta m) \ , \qquad
\hbox{where} \quad
\Delta l = -r+1, \ldots, r-1 \ , \quad
\Delta m = -s+1, \ldots, s-1 \ ,
\ee
and $\Delta l$ and $\Delta m$ only take every other value. Since
either $r$ or $s$ (or both) are even, we can only get $(l,m)$ in this
fusion product for
\be
l + \Delta l = q - l \ , \qquad
m + \Delta m = p - m \ , 
\ee
\ie\ for
\be\label{s1}
l = \frac{1}{2} \left(q - \Delta l \right) \ , \qquad
m = \frac{1}{2} \left(p - \Delta m \right) \ .
\ee
Note that in all three cases (a)--(c) these numbers are integers. 
Since $\Delta l$ ($\Delta m$) takes $r$ ($s$) values, there are
therefore $rs$ solutions; in fact, one easily checks that all
of these solutions are in fact compatible with the actual fusion rules
(\ref{fr}). Furthermore, they are pairwise identical since for
$(l,m)$ given by (\ref{s1}) 
\be
q-l = \frac{1}{2} \left(q + \Delta l \right) \ , \qquad
p-m = \frac{1}{2} \left(p + \Delta m \right) \ .
\ee
Thus there are always precisely $rs/2$ allowed torus one-point
functions, and they will precisely correspond to the $rs/2$ different
solutions of the modular differential equation.

\subsection{Null vectors at level four} \label{level4}

Given the results of the previous subsection, we shall now make a
general analysis for the torus one-point functions of the fields 
$(r,s)=(1,4)$, $(4,1)$ and $(r,s)=(2,2)$. Each of these representations
has a null-vector at level four, which is of the form
\be \label{nv}
{\cal N}_4 = \Bigl( a_1 L_{-4} + a_2 L_{-1}L_{-3} 
+ a_3 L_{-2} L_{-2} + a_4 L_{-1} L_{-1} L_{-2} 
+ a_5 L_{-1}^4 \Bigr) |h\rangle \ ,
\ee
where the parameters $a_i$ are given as 
\begin{eqnarray}
& (1,4): & a_1 = - 6 a_5 q \frac{ p^2 + 4pq + 6q^2}{p^3} \quad
             a_2 = 2 a_5 q \frac{5p + 12q}{p^2}  \quad 
             a_3 = \frac{9 a_5 q^2}{p^2}  \quad
             a_4 = - \frac{10 a_5 q}{p}  \nonumber \\
& (4,1): & a_1 = - 6 a_5 p \frac{ q^2 + 4pq + 6p^2}{q^3} \quad
             a_2 = 2 a_5 p \frac{5q + 12p}{q^2}  \quad 
             a_3 = \frac{9 a_5 p^2}{q^2}  \quad 
             a_4 = - \frac{10 a_5 p}{q}  \nonumber \\
& (2,2): & a_1= - 3 a_5 \frac{  p^2 + 2pq + q^2}{pq} \quad
             a_2=2a_5 \frac{p^2 + 3pq + q^2}{pq} \quad 
             a_3=a_5 \frac{p^4 - 2p^2 q^2 + q^4}{p^2 q^2} \nonumber \\
& &   a_4= - 2 a_5 \frac{p^2 + q^2}{pq} \ . \nonumber
\end{eqnarray}
A null vector of the form (\ref{nv}) leads to the modular
differential equation 
\be\label{deq}
\left[D_2\, D_0+\left(\frac{c+8h}{2}+3h\frac{a_1}{a_3}\right)
\frac{E_4(q)}{720} \right]\, g(q)=0 \ ,
\ee
where we have written $\chi(h;\tau)=\eta^{2h}(q)\, g(q)$ --- see
(\ref{ana1}) above. The two different solutions for $g$ have the
$T$-matrix  
\be
T=\begin{pmatrix}\exp (\pi i/6(1+\sqrt{1-144 \Delta}))& 0 \\ 
0 & \exp (\pi i/6(1-\sqrt{1-144 \Delta})) \end{pmatrix},
\ee
where
\be
\Delta =\left(\frac{c+8h}{2}+3h\frac{a_1}{a_3}\right)\frac{1}{720}\ .
\ee

\subsubsection{The case (2,2) }

If $h=h_{2,2}$, $\Delta$ is in fact independent of $p,q$ and equal to
$-\frac{5}{576}$. Moreover the $T$-matrix satisfies the 
condition for admissibility described in section~2.1. In fact, one
finds that the two solutions are given as 
\be\label{99}
g_1(q)=\frac{\eta(q^4)^2}{\eta(q)\eta(q^2)} \ , \qquad
g_2(q)=\frac{\eta(q^2)^5}{\eta(q)^3\eta(q^4)^2}\ .
\ee
Their leading behaviour is 
\be
g_i(q) = q^{s_i} \Bigl(1 + {\cal O}(q) \Bigr) \ , \qquad \hbox{with}
\quad s_1 = \frac{5}{24} \ , \quad s_2 = - \frac{1}{24} \ . 
\ee
Note that these correspond to the conformal dimensions 
\be
s_i = h_{r_i,s_i} - \frac{c}{24} - \frac{h_{2,2}}{12} \ , \quad 
\hbox{for} \quad
(r_1,s_1)=  \left(\frac{q-1}{2},\frac{p+1}{2}\right) \ , \qquad 
(r_2,s_2) = \left(\frac{q-1}{2},\frac{p-1}{2}\right) \ .
\ee
[The term $h_{2,2}/12$ comes from the $\eta$-function prefactor in the
definition of $g(q)$.] This is compatible with the allowed fusion
rules 
\be
\left(\frac{q-1}{2},\frac{p+1}{2} \right) \otimes (2,2) \supset 
\left(\frac{q+1}{2},\frac{p-1}{2} \right)
\cong \left(\frac{q-1}{2},\frac{p+1}{2}\right)
\ee
and
\be
\left(\frac{q-1}{2},\frac{p-1}{2} \right) \otimes (2,2) \supset 
\left(\frac{q+1}{2},\frac{p+1}{2} \right)
\cong \left(\frac{q-1}{2},\frac{p-1}{2}\right) \ .
\ee
Here we have assumed that $p$ and $q$ are odd; if either $p$ or $q$ is
even, then we know from the general analysis of section~3.2 that there
is a second differential equation, and that the fusion rules do not 
actually allow for any non-trivial one-point function. 

It is also not difficult to work out the corresponding representation
of the modular group and to show (for example, using the 
results of \cite{Terry3}) that it is indeed admissible.

\subsubsection{The cases of $(1,4)$ and $(4,1)$}

If $h=h_{1,4}$ the $T$-matrix takes the form
\be \label{tmat}
T=\begin{pmatrix}\exp (\pi i \frac{1+3q/p}{6})& 0 \\ 
0 & \exp (\pi i \frac{1-3q/p}{6}) \end{pmatrix}\ .
\ee
Obviously, the $(1,4)$ field only exists if $p\geq 5$ (and $q\geq 2$). 
It also follows from our general analysis that we only get 
non-trivial torus one-point functions if $p$ is odd and $q$ is 
even --- we are here in case (c). For such $p$ and $q$ the $T$-matrix 
generically fails the test of section~2.1 and therefore the
associated representation of the modular group is not admissible. 
In fact, there are only finitely many $(p,q)$ for which this is not the case; 
the simplest case is the tricritical Ising model with $p=6$ and $q=5$ that
will be discussed below in section~3.4.

To see that there are only finitely many cases where the representation
is admissible, we use the test of section~2.1. In the current context
it requires that for any $l$ coprime to $12p$, 
$l^2 (p+3q)$ must be congruent to $p\pm 3q$ (mod $12p$). 
We want to show that this is only possible if 
$p$ divides $120$. First we show that if $p$ contains as a factor
$a$, then the condition, taken mod $a$, becomes
$l^2 = \pm 1$ (mod $a$) for any $l$ coprime to $a$. 
If $a$ is a prime $a\geq 7$, we can take $l=2$ to see that this has no
solution.   
Thus $p$ can only contain the factors $2,3$ and $5$. For $a=5$,
$l=2$ is possible, but not for $a=25$; hence $p$ can only contain a single
power of $5$. Similarly, $l=2$ is possible for $a=3$, but not for 
$a=9$, and thus also only a single factor of $3$ can appear in $p$. Finally,
for $a=2$, we can take $l=3$ to conclude that at most three powers of $2$
appear in $p$. Hence $p$ must divide $2^3 \cdot 3 \cdot 5=120$, and
thus in particular $2\leq q < p \leq 120$, leading to finitely many
cases only.\footnote{We thank Terry Gannon for explaining this
argument to us.} 
\medskip

\noindent The situation is similar for $h=h_{4,1}$, for which the $T$-matrix
takes the form 
\be
T=\begin{pmatrix}\exp (\pi i \frac{1+3p/q}{6})& 0 \\ 
0 & \exp (\pi i \frac{1-3p/q}{6}) \end{pmatrix}\ .
\ee
Obviously, $q\geq 5$, and hence $p\geq 6$. Furthermore --- we are now
in case (b) --- $p$ has to be even and $q$ odd. As before, generically 
the $T$-matrix fails the test of section~2.1, except if $q=5$, where
it is satisfied for all $p$. (Note that for $q=5$, the $T$-matrix is periodic
in $p$ with period $20$.)
\smallskip

As an example where the modular representation is not admissible, let
us consider the $(7,2)$ model with central charge $c=-\frac{68}{7}$ and 
highest-weight representations
\begin{eqnarray}
& & h_{1,1}=h_{1,6}=0 \ , \qquad
h_{1,2}=h_{1,5}=-\frac{2}{7} \ , \qquad
h_{1,3}=h_{1,4}=-\frac{3}{7}\ .
\end{eqnarray}
We are interested in the $(1,4)$ field, whose $T$-matrix is (see (\ref{tmat}))
\be\label{Tsick}
T=\begin{pmatrix}\exp (\pi i \frac{13}{42})& 0 \\ 
0 & \exp (\pi i \frac{1}{42}) \end{pmatrix}\ 
\ee
and has order $N=84$. One easily checks that it fails the test of section~2.1, 
for example take $m=5$. Moreover, if we make the ansatz
\be 
g(q)=\sum_{n=0}^{\infty}a_nq^{n+s}\qquad a_0=1
\ee
in (\ref{deq}), we get the solutions
\begin{eqnarray}
\left(h_{1,2}=h_{1,5}=-\frac{2}{7}\right):&  g_{-\tfrac{2}{7}}(q)= &
q^{\frac{13}{84}}\left(1-\frac{13}{14}q-\frac{13}{49}q^2+\frac{299}{686}q^3
 -\frac{2674}{7725}q^4 +\dots\right) \ ,
\nonumber \\ 
\left(h_{1,3}=h_{1,4}=-\frac{3}{7}\right):&  g_{-\tfrac{3}{7}}(q)=&
q^{\frac{1}{84}}\left(1-\frac{4}{7}q-\frac{267}{637}q^2
+\frac{8}{343}q^3 -\frac{2236}{7203}q^4 + \dots\right)\ . \nonumber
\end{eqnarray}
Note that the  first character is taken in the $-\tfrac{2}{7}$ 
representation, since 
$-\tfrac{2}{7}-\tfrac{c}{24} - \tfrac{h_{1,4}}{12} = \tfrac{13}{84}$,
while the same analysis shows that the second is taken in the
$-\tfrac{3}{7}$ representation. The fact that  
the coefficients of the $q$-expansion are fractional numbers also suggests
that we cannot express these functions in terms of standard transcendental
functions.

Actually, the situation is even worse in that the above
two-dimensional representation of the modular group (whose
$T$-representative is given in (\ref{Tsick})) does not even have 
finite image, {\it i.e.}\ the set of $2\times 2$ matrices
corresponding to all the elements of the modular group is infinite. In
fact, all finite two-dimensional matrix groups are known --- they are
McKay's A-D-E groups --- and one can easily show that the above
representation is not one of them. (The $T$-matrix has order $84$, and
this rules out that the representation is of A- or E-type. Furthermore, the
determinant of the $T$-matrix has order $6$, and thus the group must
have a one-dimensional representation whose order is a multiple of
$6$; this rules out the D-series. We thank Terry Gannon for explaining
this to us.)

\subsection{Tricritical Ising model}

An interesting exception to these general considerations is the
tricritical Ising model with $(p,q)=(5,4)$. It has central charge 
$c=\frac{7}{10}$. The highest weight representations are 
\begin{eqnarray}
& & h_{1,1}=h_{3,4}=0\ , \quad
h_{1,2}=h_{3,3}=\frac{1}{10} \ ,\quad
h_{1,3}=h_{3,2}=\frac{3}{5} \ , \quad
h_{1,4}=h_{3,1}=\frac{3}{2} \ ,\nonumber \\
& & h_{2,1}=h_{2,4}=\frac{7}{16} \ , \quad
h_{2,2}=h_{2,3}=\frac{3}{80} \ . 
\end{eqnarray}
From our general analysis, we know that there are no torus one-point
functions for $h=\tfrac{7}{16}$ and $h=\tfrac{3}{80}$. On the other
hand, we expect precisely one torus one-point function for 
$h=\tfrac{1}{10}$, two for $h=\tfrac{3}{2}$ and three for 
$h=\tfrac{3}{5}$. 
\smallskip

\noindent In the first case, the torus one-point function is simply (see also
\cite{Mi})
\be
\chi_{3/80}(1/10,\tau) = \eta^{1/5}(q) \ .
\ee
For the case of the $\tfrac{3}{2}$ state, the differential equation
turns out to be 
\be
\left(D_{7/2} D_{3/2} -\frac{119}{3600} E_4(q) \right)
\chi(3/2;\tau) = 0 \ .
\ee
We make the ansatz
\be
\chi(3/2;\tau) = \eta^3(q) \, g(q) \ , \qquad 
g(q)=\sum_{n=0}^{\infty}a_n q^{n+s}\ , \quad a_0=1\ .
\ee
The indicial equation reads 
\begin{equation}\label{19}
s^2-\frac{s}{6}-\frac{119}{3600}=0\ ,
\end{equation}
with solutions $s_1=17/60$ and $s_2=-7/60$, corresponding to the two 
highest weight representations 
\be
\frac{3}{24}+s_1+\frac{c}{24}=\frac{7}{16} \qquad 
\frac{3}{24}+s_2+\frac{c}{24}=\frac{3}{80}\ ,
\ee
in agreement with the fusion rules
\be
\left(\frac{3}{2}\right) \otimes \left(\frac{7}{16}\right) = 
\left(\frac{7}{16}\right) \ , \qquad
\left(\frac{3}{2}\right) \otimes \left(\frac{3}{80}\right) = 
\left(\frac{3}{80}\right) \ .
\ee
The first few coefficients are explicitly given as 
\begin{eqnarray}
\left(h_{2,1}=h_{2,4}=\frac{7}{16}\right): & 
g_{\tfrac{7}{16}}(q)= & q^{\frac{17}{60}}\left(1
+ \frac{34}{7}q+ 17q^2+ 46q^3 +117q^4+ 266q^5+\cdots \right) \nonumber \\
\left(h_{2,2}=h_{2,3}=\frac{3}{80}\right): & 
g_{\tfrac{3}{80}}(q)= & q^{-\frac{7}{60}}
\left(1+ 14q+ 42q^2+ 140q^3+ 350q^4+ 840q^5+ \cdots\right) \ . \nonumber
\end{eqnarray}
It is remarkable that the coefficients are all positive integers
(after rescaling the first function by $7$). In fact, the two
functions agree precisely with the two characters of the 
$\hat{\frak g}_2$ level $k=1$ WZW model: the function $g_2$ is the
vacuum representation, while $7 g_1$ is the character corresponding to
the $7$-dimensional representation of ${\frak g}_2$. The associated
representation of the modular group is obviously admissible in this
case. 
\medskip

This leaves us with analysing the torus one-point amplitude for the
highest weight state $h=\tfrac{3}{5}$. The associated modular
differential equation takes the form
\be
\left(D_{23/5} D_{13/5} D_{3/5} 
-\frac{155}{2304} E_4(q) D_{3/5}
+\frac{25}{55296}E_6(q)\right) \chi(3/5;\tau)= 0 \ .
\ee
Writing as before $\chi_l(3/5;\tau)=\eta^{6/5} g_l(q)$ we find the 
three solutions
\begin{eqnarray}
g_{3/80}(q) & = & \frac{\eta(\tau)}{\eta(2\tau)} \\
g_{1/10}(q) & = & 
\frac{\eta(\tau)}{\eta(\tau/2)}
+e^{\pi i/24} \frac{\eta(\tau)}{\eta(\tau/2+1/2)} \\
g_{3/5}(q) & = & 
\frac{\eta(\tau)}{\eta(\tau/2)}
- e^{\pi i/24}\frac{\eta(\tau)}{\eta(\tau/2+1/2)} \ .
\end{eqnarray}
It is straightforward to determine the $S$-matrix corresponding to these 
three functions, and one finds 
\be
S=\begin{pmatrix} 0 &1/ \sqrt{2} & 1/\sqrt{2}  \\ 1/\sqrt{2} & 1/2 &-1/2 \\
1/\sqrt{2} & -1/2 & 1/2 \end{pmatrix}\ .
\ee
Since the $S$-matrix does not have a strictly positive row or column,
it does not describe the $S$-matrix of a conformal field theory; in
particular, there is no Verlinde formula associated to it. This also
ties in with the fact that $g_{3/80}(q)$ has (integer) coefficients of
both signs, and hence cannot be interpreted as a character. However,
the associated representation of the modular group is certainly
admissible.

\section{Conclusions}
\setcounter{equation}{0}

In this paper we have determined torus one-point functions 
by solving the associated modular differential equations. The 
underlying method is very general and applies 
in principle to any rational conformal field theory. We 
have exemplified it for the case of the Virasoro minimal models.
For some low-lying cases we could give explicit formulae in terms
of well-known functions. In particular, this was possible for
all torus one-point functions of the Yang-Lee, the Ising and the
tricritical Ising model. (Some of these results had been found
before in \cite{Francesco,Mi}.) However, in general the solutions
are more complicated and cannot be expressed in terms of
standard transcendental functions (as we also exhibited).
Probably the resulting functions are still fairly special since
they arise in very special conformal field theories; it would
be very interesting to understand their structure better.

\section*{Acknowledgements}

This research has been partially supported by 
the Swiss National Science Foundation and the Marie Curie network
`Constituents, Fundamental Forces and Symmetries of the Universe'
(MRTN-CT-2004-005104). We thank Terry Gannon for many useful 
discussions and correspondences. This paper is largely based on the 
Diploma thesis of S.L.

\appendix
\renewcommand{\theequation}{\Alph{section}.\arabic{equation}}

\section{Vertex operator algebras}\label{voa}
\setcounter{equation}{0}

Let us begin by collecting our conventions. The vacuum representation
of a (chiral) conformal field theory describes a meromorphic conformal
field theory \cite{pg}. In mathematics, this structure is usually
called a vertex operator algebra (see for example \cite{FLM,Kac} for a
more detailed introduction). A vertex operator algebra is a vector
space $V = \bigoplus_{n=0}^\infty V_n$ of states, graded by the
conformal dimension. Each element in $V$ of grade $h$ defines a linear 
map on $V$ via
\be\label{modex}
a \mapsto  V(a,z) = \sum_{n\in \mathbb{Z}} 
a_n\, z^{-n-h}\qquad (a_n \in {\rm End}\ V)\ .
\ee
In this paper we follow the usual physicists' convention for the 
numberings of the modes; this differs by a shift by $h-1$
from the standard 
mathematical convention that is also, for example, used in
\cite{Zhu}. 
We also use sometimes (as in \cite{Zhu}) the symbol
\be
o(a) = a_0 \ . 
\ee
Since much of our analysis is concerned with torus amplitudes
it will be convenient to work with the modes that naturally appear
on the torus; they can be obtained via a conformal transformation
from the modes on the sphere. More specifically, we define
(see section 4.2 of \cite{Zhu}) 
\be\label{A4}
V[a,z] = e^{2\pi i z h_a}\, V(a,e^{2\pi i z}-1) 
= \sum_n a_{[n]}\, z^{-n-h} \ . 
\ee
The explicit relation is then
\be
a_{[m]} = (2\pi i)^{-m-h_a}\sum_{j\geq m} c(h_a,j+h-1,m+h-1) \,a_j \ , 
\ee
where 
\be
(\log(1+z))^m (1+z)^{h_a-1} = \sum_{j\geq m} c(h_a,j,m) \, z^j \ . 
\ee
This defines a new vertex operator algebra with a new Virasoro
tensor  whose modes $L_{[n]}$ are given by
\be\label{A14}
L_{[n]}= (2\pi i)^{-n}\sum_{j\geq n+1} c(2,j,n+1) L_{j-1} -
(2\pi i)^2\frac{c}{24}\delta_{n,-2} \ .
\ee
The appearance of the correction term for $n=-2$ is due to the fact that
$L$ is only quasiprimary, rather than primary. Since the two descriptions 
are related by a conformal transformation to one another, the new modes
$S_{[n]}$ satisfy the same commutation relations as the original modes
$S_n$. In particular, the modes $L_{[n]}$ satisfy a Virasoro algebra with
the same central charge as the modes $L_n$. 

\section{Eisenstein series and modular covariant derivative}
\setcounter{equation}{0}
\label{eisenstein}

The Eisenstein series $G_{2k}(q)$, $q=e^{2\pi i\tau}$, are defined by
\begin{eqnarray}
G_{2k}(q)&=&\sum_{(m,n)\neq(0,0)}\frac{1}{(m\tau+n)^{2k}}\ ,
\quad k\geq 2\ , \\ 
G_2(q)&=&\frac{\pi^2}{3}+\sum_{m\in\mathbb{Z}\backslash \{0\}} 
\sum_{n\in \mathbb{Z}} \frac{1}{(m\tau + n)^2}\ .
\end{eqnarray}
For $k\geq 2$, the Eisenstein series are modular forms of weight $2k$,
that is 
\begin{equation}
G_{2k}\left(\frac{a\tau +b}{c\tau +d}\right)
=(c\tau +d)^{2k}\, G_{2k}(\tau)\ ,
\end{equation}
whereas $G_2$ transforms as
\begin{equation}
G_{2}\left(\frac{a\tau +b}{c\tau +d}\right)
=(c\tau + d)^2 G_2(\tau) -2 \pi i c(c\tau +d)\ .
\end{equation}

This modular anomaly of $G_2$ can be used to define a modular
covariant derivative. Suppose $f(q)$ is a modular form of weight $s$,
then $D_sf(q)$ is a modular form of weight $s+2$, where 
\be
D_s=q\frac{d}{dq}-\frac{s}{4 \pi ^2}G_2(q)\ .
\ee
For $k\geq 4$ the space of modular forms of weight $k$ has a basis
(see for example chapter 4 in \cite{Forms}) 
\be
\{E_4(q)^m E_6(q)^n|4m+6n=k,\,\,m,n\geq 0\}\ ,
\ee
where $E_n=\frac{G_n}{2\zeta(n)}$ denotes the normalised Eisenstein
series, such that the constant term in the power series expansion is 
$1$. In particular all higher $G_{2k}$ can be written as polynomials
in $G_4$, $G_6$. The normalised Eisenstein series are given by  
\begin{eqnarray}
E_2(q)&=&1-24 q-72 q^2-96 q^3-168 q^4-144q^5-288q^6-\dots \nonumber \\
E_4(q)&=&1+240 q+2160 q^2+6720 q^3+ 17520 q^4 + 30240 q^5 
+ 60480 q^6 +\dots\\
E_6(q)&=&1-504q - 16632q^2-122976q^3 - 532728 q^4 - 1575504 q^5\nonumber\\
&& - 4058208q^6 - \dots\ , \nonumber
\end{eqnarray}
and the relation between $G_n$ and $E_n$ for the first few values reads
\begin{equation} 
G_2(q)=-\frac{(2\pi i)^2}{12} E_2(q)\ ,\quad
G_4(q)=\frac{(2\pi i)^4}{720}E_4(q)\ ,\quad
G_6(q)=-\frac{(2\pi i)^6}{30240}E_6(q)\ . 
\end{equation}

\subsection{Differential operators}\label{diffop}
The explicit formulae for the differential operators $P_r^h(D)$ are
given by 
\begin{eqnarray*}
P_1^{(h)}(D)&=&(2\pi i)^2 D^{(1,h)}\\
P_2^{(h)}(D)&=&(2\pi i)^4D^{(2,h)} +\frac{c+8h}{2} G_4(q)\\
P_3^{(h)}(D)&=&(2\pi i)^6 D^{(3,h)}
+ \left(8+\frac{3(c+8h)}{2}\right)G_4(q)
(2\pi i)^2D^{(1,h)}+10(c+8h)G_6(q)\\
P_4^{(h)}(D)&=&(2\pi i)^8 D^{(4,h)} 
+ (32+3(c+8h))G_4(q) (2\pi i)^4 D^{(2,h)}\\ 
&&+(160+40(c+8h))G_6(q) (2\pi i)^2 D^{(1,h)}
+\big(108(c+8h)+\frac{3}{4}(c+8h)^2\big)G_4(q)^2\ ,
\end{eqnarray*}
where $c$ is the central charge.

\section{Test of admissibility}\label{proof}
\setcounter{equation}{0}

In this appendix we explain the simple criterion of admissibility
(that is due to Terry Gannon).\footnote{We thank Terry Gannon
for communicating also the proof to us.} Let $\rho$ be a representation of 
$\mbox{SL}_2(\mathbb{Z})$, such that the matrix $T$ defined in 
(\ref{Tdef}) is diagonal. Let $t_1,\dots,t_m$ be the diagonal elements of 
$T$. Suppose that the kernel of $\rho$ contains $\Gamma(N)$. 
Then for all integers $\ell$ coprime to $N$, $t_1^{\ell^2},\dots,t_m^{\ell^2}$ 
is identical with $t_1,\dots,t_m$, as multi-sets (\ie\ the order may have 
changed, but the multiplicities must be identical). 
\medskip

\noindent {\bf Proof:}
Since $\Gamma(N)$ is in the kernel of $\rho$, $\rho$ is well-defined
as a representation of the quotient
$\mbox{SL}_2(\mathbb{Z})/\Gamma(N)$. But this quotient is just
$\mbox{SL}_2(\mathbb{Z}_N)$, \ie\ the set of $2\times 2$ matrices with 
entries in the ring $\mathbb{Z}_N:=\mathbb{Z}/N\mathbb{Z}$ and
determinant $\equiv 1$ (mod $N$). Now, if $\ell$ is coprime to $N$,
then $\ell$ has a multiplicative inverse mod $N$, \ie\ there is an
integer, which we shall call $\ell^{-1}$, with the property that
$\ell\ell^{-1}\equiv 1$ (mod $N$). This means that for any $\ell$
coprime to $N$, the matrix 
\be
D_\ell:=\begin{pmatrix} \ell^{-1}&0 \\ 0& \ell \end{pmatrix}
\ee
lies in $\mbox{SL}_2(\mathbb{Z}_N)$, and has inverse
\be
D_\ell^{-1}=\begin{pmatrix} \ell & 0\\ 0&\ell^{-1}\end{pmatrix}.
\ee
Now we use that in $\mbox{SL}_2(\mathbb{Z}_N)$ we have the identity
\be
\begin{pmatrix} \ell & 0\\ 0&\ell^{-1} \end{pmatrix} 
\begin{pmatrix} 1 &1\\0&1\end{pmatrix} 
\begin{pmatrix} \ell^{-1}&0\\ 0&\ell\end{pmatrix} = 
\begin{pmatrix}1&\ell^2\\0&1\end{pmatrix}\ .
\ee
But any $M$ in $\mbox{SL}_2(\mathbb{Z}_N)$ can be lifted to
$\mbox{SL}_2(\mathbb{Z})$, \ie\ we can find a matrix 
$M'\in\mbox{SL}_2(\mathbb{Z})$, such that $M' \equiv M$ (mod
$N$). So this means 
\be
\rho(D'_\ell)^{-1}T\rho(D'_\ell)=T^{\ell^2}\ ,
\ee
\ie\ the matrices $T$ and $T^{\ell^2}$ are conjugate to each
other. Since $T$ and $T^{\ell^2}$ are diagonal, this is equivalent to
the statement that their diagonal elements are identical (as
multi-sets). 
\bigskip

It is now straightforward to deduce the following corollary from
this statement. Let $\rho$ be a representation
of $\mbox{SL}_2(\mathbb{Z})$, such that the $T$-matrix is
diagonal. Let $t_1,\dots,t_m$ be the eigenvalues of $T$ and let $N$ be
the order of $T$, \ie\ $T^N={\bf 1}$. Suppose there is an integer $\ell$ 
coprime to $N$, such that $t_1^{\ell^2},\dots,t_m^{\ell^2}$ is not
identical with $t_1,\dots,t_m$, as multi-sets. Then $\rho$ does not
contain any congruence subgroup $\Gamma(N')$ in its kernel. 
In particular, it therefore does not satisfy condition 1 of 
section 2.1, and hence is not admissible.
\medskip

\noindent {\bf Proof:} Suppose for contradiction that
$\rho$ contains $\Gamma(N')$ in its kernel, for some $N'$. Then the
order $N$ of $T$ must divide $N'$. Let $\ell$ be the integer coprime
to $N$ with the property stated in the previous paragraph. Lift $\ell$ to an
integer $\ell '$ coprime to $N'$, \ie\ find an integer $\ell '$ such
that $\ell '$ is coprime to $N'$, and $\ell '\equiv\ell$ (mod
$N$). Such an $\ell '$ exists by the Chinese remainder theorem. Then
$T^{\ell '^2}=T^{\ell^2}$ since $T$ has order $N$. So this $\ell '$
contradicts the previous statement.


\begin{thebibliography}{99}

\bibitem{Eguchi:1986sb}
T.~Eguchi and H.~Ooguri,
{\it Conformal and current algebras on a general Riemann surface},
Nucl.\ Phys.\  B {\bf 282} (1987) 308.

\bibitem{Anderson:1987ge}
G.~Anderson and G.W.~Moore,
{\it Rationality in conformal field theory},
Commun.\ Math.\ Phys.\  {\bf 117} (1988) 441.

\bibitem{MMS} 
S.D.~Mathur, S.~Mukhi and A.~Sen, 
{\it  On the classification of rational conformal field theories}, 
Phys.\ Lett.\ B {\bf 213} (1988) 303.

\bibitem{MMS1} 
S.D.~Mathur, S.~Mukhi and A.~Sen,
{\it  Reconstruction of conformal field theories from modular geometry
on the torus}, Nucl.\ Phys.\ B {\bf 318} (1989) 483. 

\bibitem{Kiritsis:1989kc}
E.B.~Kiritsis,
{\it Analytic aspects of rational conformal field theories}, 
Nucl.\ Phys.\  B {\bf 329} (1990) 591.

\bibitem{Eholzer} 
W.~Eholzer, 
{\it On the classification of modular fusion algebras},   
Commun.\ Math.\ Phys.\ {\bf 172} (1995) 623 {\tt [hep-th/9408160]};
Ph.D. thesis (Bonn 1995), {\tt hep-th/9502160}.  

\bibitem{ES}
W.~Eholzer and N.-P.~Skoruppa, 
{\it Modular invariance and uniqueness of conformal characters}, 
Commun.\ Math.\ Phys.\ {\bf 174} (1995) 117 {\tt [hep-th/9407074]}; 
{\it Conformal characters and theta series}, 
Lett.\ Math.\ Phys.\ {\bf 35} (1995) 197 {\tt [hep-th/9410077]}. 

\bibitem{Zhu} 
Y.~Zhu, 
\textit{Modular invariance of characters of vertex operator algebras},
J.\ Amer.\ Math.\ Soc \textbf{9} (1996) 237. 

\bibitem{Nahm:1991ie}
W.~Nahm,
{\it A proof of modular invariance},
Int.\ J.\ Mod.\ Phys.\  A {\bf 6} (1991) 2837.

\bibitem{Gaberdiel:2007ve}
M.R.~Gaberdiel,
\textit{Constraints on extremal self-dual CFTs},
JHEP {\bf 0711} (2007) 087
{\tt [arXiv:0707.4073 [hep-th]]}.

\bibitem{GaKe}
M.R.~Gaberdiel and C.A.~Keller,
\textit{Modular differential equations and null vectors},
JHEP {\bf 0809} (2008) 079 
{\tt [arXiv:0804.0489 [hep-th]]}.

\bibitem{Flohr:2005cm}
M.~Flohr and M.R.~Gaberdiel,
{\it Logarithmic torus amplitudes},
J.\ Phys.\ A  {\bf 39} (2006) 1955 
{\tt  [hep-th/0509075]}.

\bibitem{Mi}
M.~Miyamoto,
\textit{ Intertwining operators and modular invariance},
{\tt math/0010180}. 

\bibitem{Francesco} 
P.~Di Francesco, H.~Saleur and J.-B.~Zuber, 
\textit{Critical Ising correlation functions in the plane 
and on the torus}, 
Nucl.\ Phys.\ B \textbf{290} (1987) 527.

\bibitem{Bernard:1987df}
D.~Bernard,
\textit{On the Wess-Zumino-Witten models on the torus},
Nucl.\ Phys.\  B \textbf{303} (1988) 77.

\bibitem{Felder:1994gk}
G.~Felder and C.~Weiczerkowski,
{\it Conformal blocks on elliptic curves and the
Knizh\-nik-Zamolodchikov-Bernard equations},
Commun.\ Math.\ Phys.\  \textbf{176} (1996) 133
{\tt [hep-th/9411004]}.

\bibitem{Huang:2003yq}
Y.Z.~Huang,
\textit{ Differential equations and conformal field theories},
{\tt math/0303269.}

\bibitem{Felder} 
G.~Felder and R.~Silvotti, 
\textit{Modular covariance of minimal model correlation functions},
Commun.\ Math.\ Phys.\ {\bf 123} (1989) 1.

\bibitem{Terry1} 
P.~Bantay and T.~Gannon, 
\textit{Conformal characters and the modular representation},
JHEP {\bf 0602} (2006) 005 {\tt [hep-th/0512011]}.

\bibitem{Terry2}
P.~Bantay and T.~Gannon,
{\it Vector-valued modular functions for the modular group and the
hypergeometric equation},  
{\tt arXiv:0705.2467 [math.NT]}.

\bibitem{Abe}
T.~Abe, G.~Buhl and C.~Dong,
\textit{Rationality, regularity and $C_2$ co-finiteness},
Trans.\ Amer.\ Math.\ Soc.\ {\bf 356} (2004) 3391
{\tt [math/0204021]}. 

\bibitem{Terry3} 
A.~Coste and T.~Gannon, 
\textit{Congruence subgroups and rational conformal field theory},
{\tt math/9909080}. 

\bibitem{Lang}
S.~Lang,
\textit{Elliptic Functions}, 
(Springer, 1987).

\bibitem{DMS}
P.~Di Francesco, P.~Mathieu and D.~S\'en\'echal, 
\textit{Conformal Field Theory}, (Springer, 1997).

\bibitem{pg}
P.~Goddard, 
{\it Meromorphic conformal field theory}, in:
{\it Infinite dimensional Lie algebras and Lie groups: Proceedings  
of the CIRM Luminy Conference, 1988} (World Scientific, Singapore,
1989) 556.

\bibitem{FLM}
I.~Frenkel, J.~Lepowsky and A.~Meurman, 
{\it Vertex operator algebras and the Monster}, (Academic Press,
1988). 

\bibitem{Kac}
V.G.~Kac,
{\it Vertex algebras for beginners}, (AMS, 1998). 

\bibitem{Forms} 
T.~Miyake, 
\textit{Modular Forms}, (Springer, 2006).

\bibitem{Apostol} 
T.M.~Apostol, 
\textit{Modular Functions and Dirichlet Series in Number Theory}, 
(Springer, 1997). 


\end{thebibliography}
\end{document}